\documentclass[useAMS,usenatbib,usegraphicx]{mn2e}

\title[Very small-scale clustering of quasars]
{Very small-scale clustering of quasars from a complete quasar lens survey}

\author[I. Kayo and M.~Oguri]
{Issha Kayo$^{1}$\thanks{kayo@ph.sci.toho-u.ac.jp} and Masamune Oguri$^{2}$\\
$^1$Department of Physics, Toho University, 2-2-1 Miyama, Funabashi,
  Chiba 274-8510, Japan.\\ 
$^2$Kavli Institute for the Physics and Mathematics of the Universe
  (Kavli IPMU, WPI), Todai Institutes for Advanced Study, \\
  University of Tokyo, Kashiwa, Chiba 277-8583, Japan.\\ 
} 

\begin{document}

\date{\today}


\pagerange{\pageref{firstpage}--\pageref{lastpage}} \pubyear{}

\maketitle

\label{firstpage}

\begin{abstract}
We measure the small-scale (comoving separation 
$10h^{-1}{\rm kpc}\la r_p \la 200 h^{-1}{\rm kpc}$) 
two-point correlation function of quasars using a sample of 26
spectroscopically confirmed binary quasars at $0.6<z<2.2$ from the
Sloan Digital Sky Survey Quasar Lens Search (SQLS). Thanks to careful
candidate selections and extensive follow-up observations of the SQLS,
which is aimed at constructing a complete quasar lens sample, our
sample of binary quasars is also expected to be nearly complete within
a specified range of angular separations and redshifts. The measured
small-scale correlation function rises steeply toward smaller scales,
which is consistent with earlier studies based on incomplete or
smaller binary quasar samples. We find that the quasar correlation
function can be fitted by a power-law reasonably well over 4 order of
magnitudes, with the best-fit slope of $\xi(r)\propto r^{-1.92}$. We
interpret the measured correlation function within the framework of
the Halo Occupation Distribution (HOD). We propose a simple model
which assumes a constant fraction of quasars that appear as satellites
in dark matter haloes, and find that measured small-scale clustering
signals constrain the satellite fraction to 
$f_{\rm sat}=0.054_{-0.016}^{+0.017}$ for a singular isothermal sphere
number density profile of satellites. We note that the HOD modelling
appears to under-predict clustering signals at the smallest separations
of $r_p\sim 10h^{-1}{\rm kpc}$ unless we assume very steep number
density profiles (such as an NFW profile with the concentration
parameter $c_{\rm vir}\ga 30$), which may be suggestive of enhanced
quasar activities by direct interactions.
\end{abstract}


\section{Introduction}
\label{sec:intro}

Quasars are luminous active galactic nuclei (AGNs) powered by gas
accretion onto supermassive black holes. The co-evolution of quasars
and galaxies, suggested by the scaling relation between the mass of
the supermassive black holes and properties of their host spheroids 
\citep{magorrian98,ferrarese00,gebhardt00,tremaine02}, implies that
studies of evolution and environment of quasars are a key to
understanding evolution of galaxies. A number of studies have argued
that quasar activity is triggered by galaxy mergers 
\citep[e.g.,][]{dimatteo05,hopkins08}. There are several observational
hints for this, including the morphology of binary quasar host 
galaxies \citep{green10} and the small-scale environment 
\citep{serber06,strand08}, although it is still not clear how and to
what extent mergers are crucial for triggering quasars 
\citep[e.g.,][]{coil07,padmanabhan09,green11}. We note that possible
enhanced activities due to close interactions have also been argued 
for moderate luminosity AGNs
\citep[e.g.,][]{koss10,silverman11,ellison11,liu12}. 

In particular, clustering measurements of quasars provide an important
means of studying the evolution and environment of quasars. Substantial 
progress have been made thanks to large-area spectroscopic surveys
such as Two-degree Field (2dF) and Sloan Digital Sky Survey
(SDSS). Large-scale quasar auto-correlation functions measured in
these surveys 
\citep{porciani04,croom05,porciani06,myers06,shen07,shen09,ross09,white12}
indicate that the clustering bias of quasars is relatively high and
increases rapidly with redshifts, which suggests that quasars live in
massive ($M_{\rm vir}\ga 3\times 10^{12}h^{-1}M_\odot$) dark matter
haloes. Yet, small-scale ($\la 100 h^{-1}{\rm kpc}$) clustering
measurements, which are crucial for testing the suggested
merger-driven quasar activities, have been difficult in these surveys,
because of the finite size of fibres used for the multi-object
spectroscopy which corresponds to $\sim 1'$ on the sky. 

While the measurement of small-scale clustering is challenging, 
\citet{hennawi06} compiled a sample of binary quasars obtained from
follow-up spectroscopy of SDSS data, and measured the projected
correlation function down to the comoving separation of 
$\sim 20 h^{-1}{\rm kpc}$ \citep[see also][for similar results at
higher redshifts]{hennawi10,shen10}. They found that the measured
small-scale clustering is stronger than the power-law extrapolation of
the large-scale clustering measurement by \citet{porciani04}. This 
``excess'' of small-scale clustering has been interpreted as enhanced
quasar activities by direct interactions. 

However, one of the potential problems about the analysis of
\citet{hennawi06} is that the follow-up observations were not complete
and therefore they had to introduce large corrections of
incompleteness. The uncertainty in the incompleteness factor therefore
directly translates into the uncertainty of the measured small-scale
clustering. Indeed, \citet{myers07,myers08} constructed more complete
sample of binary quasars from follow-up observations of the SDSS
photometric quasar catalogue for a limited range of angular
separations, and claimed that the small-scale excess is not as strong
as \citet{hennawi06} found. 

In this paper, we measure the very small-scale clustering of
low-redshift ($0.6<z<2.2$) quasars from the SDSS. We take advantage of
a binary quasar sample from the SDSS Quasar Lens Search
\citep[SQLS;][]{oguri06,oguri08,oguri12,inada08,inada10,inada12},
which is a large survey of gravitationally lensed quasars among
spectroscopically confirmed SDSS quasars. While the main focus of SQLS
is to locate lensed quasars, it has also discovered a large number of
binary quasars with the angular separation $\theta<20''$. Thanks to the
careful candidate selection, which is designed to achieve very high
completeness of the lens candidate selection, and extensive follow-up
observations for constructing a complete and robust sample of lensed
quasars, the binary quasar sample from the SQLS is also expected to be
nearly complete and uniform within a specified range of the redshift,
magnitude, and  angular separation. We use this binary quasar sample
to make an accurate measurement of the quasar correlation function 
down to $\sim 10 h^{-1}{\rm kpc}$, which is small enough for
discussing the effect of dynamical interaction and mergers on quasars. 

The structure of this paper is as follows. We describe the binary
quasar sample used for the analysis in Section~\ref{sec:data}. 
Clustering measurements are given in
Section~\ref{sec:clusteringAnalysis}. We interpret the measured
correlation function in terms of the so-called halo model in
Section~\ref{sec:halomodel}. Finally we give conclusion in
Section~\ref{sec:summary}. 
We employ a flat $\Lambda$-dominated cosmology with $\Omega_m=0.27$,
$\Omega_\Lambda=0.73$, $\sigma_8=0.8$ and the Hubble constant of $H_0=100
h{\rm \,km\,s^{-1}\,Mpc^{-1}}$. We also use a different set of
cosmological parameters with $\Omega_m=0.3$, $\Omega_\Lambda=0.7$ 
and $H_0=70{\rm \,km\,s^{-1}\,Mpc^{-1}}$ when we handle the luminosity
function of \citet{croom09}. 

\section{Data}
\label{sec:data}

\begin{table*}
 \caption{Binary quasar sample from the SQLS
 \citep{inada08,inada10,inada12}. The typical error on the velocity
 measurement is a few hundreds ${\rm km\,s^{-1}}$.
\label{tab:binary}}   
 \begin{tabular}{@{}lcccccccccc}
  \hline
   Name
   & R.A.(A)
   & Decl.(A)
   & R.A.(B)
   & Decl.(B)
   & $z_A$
   & $z_B$
   & $i_A$$^a$
   & $i_B$$^a$
   & $\Delta\theta$
   & $|\Delta v|$ \\
   & (J2000)
   & (J2000)
   & (J2000)
   & (J2000)
   &
   &
   & (mag) 
   & (mag) 
   & (arcsec) 
   & (${\rm km\,s^{-1}}$) \\
 \hline
SDSSJ0740+2926$^b$& 07 40 13.44 & +29 26 48.3 & 07 40 13.42 & +29 26 45.7 & 0.978 & 0.980 & 18.41 & 19.66 &  2.6 &  230 \\
SDSSJ0847$-$0013 & 08 47 10.41 & $-$00 13 02.7 & ---     & ---         & 0.626 & 0.627 & 18.61 & 19.04 &  1.0 &  190 \\
SDSSJ0909+5801 & 09 09 55.55 & +58 01 43.3 & 09 09 56.50 & +58 01 40.5 & 1.712 & 1.712 & 18.96 & 20.17 &  8.1 &    0 \\
SDSSJ0918+2435 & 09 18 08.86 & +24 35 50.1 & 09 18 09.07 & +24 36 04.0 & 1.218 & 1.223 & 18.52 & 19.60 & 14.2 &  680 \\
SDSSJ0942+2310 & 09 42 34.98 & +23 10 31.2 & 09 42 35.04 & +23 10 28.9 & 1.833 & 1.833 & 18.99 & 19.70 &  2.5 &    0 \\
SDSSJ1000+5406 & 10 00 34.18 & +54 06 28.6 & 10 00 34.86 & +54 06 41.5 & 1.212 & 1.215 & 18.65 & 19.14 & 14.2 &  430 \\
SDSSJ1008+0351 & 10 08 59.55 & +03 51 04.4 & ---         & ---         & 1.745 & 1.740 & 19.10 & 20.28 &  1.1 &  550 \\
SDSSJ1012+3650 & 10 12 11.30 & +36 50 30.7 & 10 12 11.07 & +36 50 14.4 & 1.678 & 1.678 & 18.81 & 20.01 & 16.6 &    0 \\
SDSSJ1035+0752$^b$ & 10 35 19.37 & +07 52 58.0 & 10 35 19.23 & +07 52 56.4 & 1.216 & 1.218 & 19.03 & 20.11 &  2.7 &  270 \\
SDSSJ1120+6711$^c$ & 11 20 12.11 & +67 11 15.9 & ---         & ---         & 1.495 & 1.495 & 18.47 & 19.55 &  1.5 &   50 \\
SDSSJ1216+4957 & 12 16 47.22 & +49 57 20.4 & 12 16 47.62 & +49 57 10.6 & 1.200 & 1.195 & 18.34 & 19.55 & 10.5 &  680 \\
SDSSJ1250+1741 & 12 50 22.32 & +17 41 44.5 & 12 50 22.32 & +17 41 44.5 & 1.246 & 1.241 & 19.06 & 18.63 & 13.9 &  650 \\
SDSSJ1254+6104$^b$ & 12 54 21.98 & +61 04 22.0 & 12 54 20.52 & +61 04 36.0 & 2.051 & 2.041 & 18.91 & 19.27 & 17.6 & 1010 \\
SDSSJ1358+2326 & 13 58 09.87 & +23 26 10.1 & 13 58 10.68 & +23 26 04.5 & 1.555 & 1.543 & 18.92 & 19.93 & 12.5 & 1400 \\
SDSSJ1400+2323 & 14 00 12.28 & +23 23 46.7 & 14 00 12.86 & +23 23 51.9 & 1.877 & 1.867 & 18.34 & 19.27 &  9.5 & 1040 \\
SDSSJ1430+0714$^d$ & 14 30 02.88 & +07 14 11.3 & 14 30 02.66 & +07 14 15.6 & 1.246 & 1.261 & 19.01 & 19.68 &  5.4 & 1990 \\
SDSSJ1433+1450 & 14 33 50.94 & +14 50 08.2 & 14 33 51.09 & +14 50 05.6 & 1.506 & 1.506 & 18.82 & 19.19 &  3.3 &    0 \\
SDSSJ1511+3357 & 15 11 09.85 & +33 57 01.7 & ---         & ---         & 0.799 & 0.799 & 18.94 & 19.63 &  1.1 &   80 \\
SDSSJ1518+2959 & 15 18 23.06 & +29 59 25.5 & 15 18 23.43 & +29 59 27.6 & 1.249 & 1.256 & 18.86 & 19.88 &  5.3 &  900 \\
SDSSJ1539+3020 & 15 39 37.74 & +30 20 23.7 & 15 39 37.10 & +30 20 17.0 & 1.644 & 1.648 & 18.67 & 19.73 & 10.8 &  450 \\
SDSSJ1552+0456 & 15 52 18.09 & +04 56 35.3 & 15 52 17.94 & +04 56 46.8 & 1.567 & 1.567 & 18.20 & 18.62 & 11.7 &    0 \\
SDSSJ1552+3009 & 15 52 25.63 & +30 09 02.1 & ---         & ---         & 0.750 & 0.750 & 18.86 & 19.43 &  1.2 &    0 \\
SDSSJ1606+2900$^d$ & 16 06 02.81 & +29 00 48.7 & 16 06 03.02 & +29 00 50.9 & 0.769 & 0.769 & 18.31 & 18.38 &  3.5 &    0 \\
SDSSJ1635+2052 & 16 35 20.05 & +20 52 25.2 & 16 35 19.51 & +20 52 13.9 & 1.775 & 1.775 & 19.03 & 20.07 & 13.6 &   90 \\
SDSSJ1655+2605 & 16 55 02.01 & +26 05 16.5 & 16 55 01.32 & +26 05 17.5 & 1.890 & 1.879 & 17.63 & 17.77 &  9.6 & 1140 \\ 
SDSSJ2111+1050 & 21 11 02.61 & +10 50 38.4 & 21 11 02.41 & +10 50 47.6 & 1.897 & 1.897 & 18.87 & 19.02 &  9.7 &  120 \\
 \hline
 \end{tabular}
\flushleft{$^a$ $i$-band PSF magnitudes with Galactic extinction correction.}
\flushleft{$^b$ Originally reported in \citet{hennawi06}.}
\flushleft{$^c$ Originally reported in \citet{pindor06},
and also included in \citet{hennawi06}.
}
\flushleft{$^d$ Originally reported in \citet{myers08}.}
\end{table*}

We use the binary quasar sample in the SQLS DR7 source quasar catalogue
\citep{inada12}.  The sample is based on the SDSS Data Release 7 (DR7)
quasar catalogue \citep{schneider10}. Among the original $\sim 100,000$
quasars in the DR7 quasar catalogue, the SQLS selects source quasars
in the redshift range of $0.6<z<2.2$ and the Galactic extinction
corrected $i$-band magnitude range of $i<19.1$ to define the source
quasar catalogue. Quasars in poor seeing fields 
({\tt PSF\_WIDTH}$>1\farcs8$ in $i$-band) are also excluded because
the selection of small-separation lenses is inefficient in these 
fields. The lens sample which is appropriate for various statistical
studies is constructed within this sample. The number of quasars after
the cut is 50,836, which will be the main sample of quasars for our
small-scale clustering measurements. 

The SQLS identifies lens candidates using the so-called morphology and
colour algorithms, which are designed to select small- ($\la 2\farcs5$)
and large-separation ($\ga 2\farcs5$) lens candidates, respectively.
The lens candidates are then observed in optical imaging,
near-infrared imaging, and optical spectroscopy to check whether they
are true gravitational lens systems or not. After careful examinations 
of follow-up observation results, a complete sample of lensed quasars
is constructed in the image separation range of $1''<\theta<20''$ and
the $i$-band magnitude difference $|\Delta i|<1.25$. Simulations of
SDSS images were conducted to confirm that the selection algorithm is
almost complete in these ranges of the image separation and magnitude
differences. Interested readers are referred to \citet{oguri06} and
\citet{inada08} for details of the lens candidate selection algorithms
and follow-up procedures. \citet{inada12} presented 52 quasar lens
contained in the original SDSS DR7 quasar catalogue, 26 of which
constitute a statistical lens sample selected from the 50,836 source
quasar sample.

On the course of the lens survey, 81 spectroscopically confirmed
quasar pairs that are not gravitational lensing were also identified
\citep{inada12}. 26 of them are physically associated binary quasars,
where we adopted the velocity difference of $|\Delta v|<v_{\rm
  max}=2000{\rm km\,s^{-1}}$ for defining binary quasars following the
criteria used in \citet{hennawi06}. We summarize the 26 binary quasars
in Table~\ref{tab:binary}, which have been reported in a series of the
SQLS lens sample papers \citep{inada08,inada10,inada12}. In this
paper, we adopt this binary quasar sample to study the small-scale
quasar clustering.

Table~\ref{tab:binary} suggests that our binary quasar sample is
significantly improved over the sample used by \citet{hennawi06}, as
only 4 binary quasars out of the 26 binary quasars were included in
the analysis of \citet{hennawi06}. There are two main reasons for
this. One is that \citet{hennawi06} used only $\sim 50$\% 
of the DR7 quasar sample, simply because the SDSS was not completed at
that time. The other is that the binary quasar sample of
\citet{hennawi06} was incomplete, with the estimated completeness of
$\sim 0.3-0.5$ at $\theta<20''$, because only a part of binary
quasar candidates have received follow-up observations.
These two facts explain the small
overlap between our binary quasar sample and that of \citet{hennawi06}.

\begin{figure}
 \includegraphics[width=84mm]{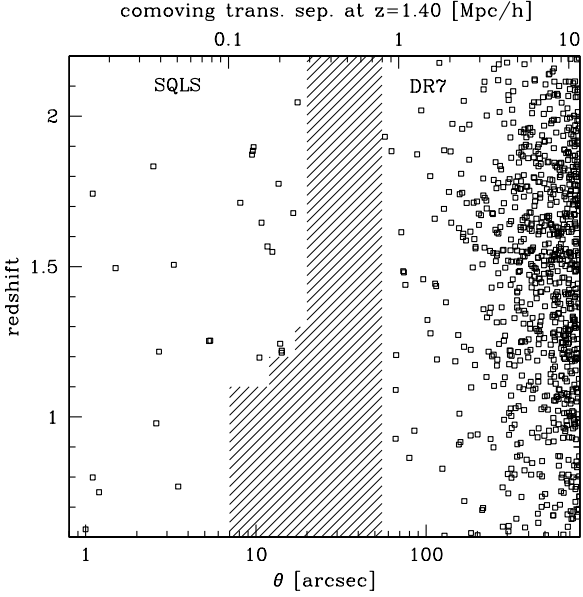}
 \caption{Angular separation and redshift distribution of our quasar
 pair sample. The points on the left-hand side of the shaded exclusion
 region are from SQLS binary catalogue (see also
 Table~\ref{tab:binary}), and those on the right-hand side are from
 the SDSS DR7 quasar catalogue. We also label the comoving transverse 
 separation at redshift of $z=1.40$ as a reference.}
 \label{fig:pairs}
\end{figure}

 We also measure clustering amplitudes of quasar pairs in the original
 SDSS DR7 quasar catalogue with separation larger than $55''$,
 which corresponds to the fibre collision limit of the SDSS
 spectrograph. The redshift and magnitude ranges we consider are same
 as the SQLS, $0.6<z<2.2$ and $i<19.1$ with the number of quasars
 of 51,283. Unlike \citet{hennawi06}, we ignore pairs with angular
 separations $20''<\theta<55''$ because the completeness in this
 angular separation range is low and uncertain. Figure~\ref{fig:pairs}
 plots the separation and redshift of pairs we consider. The shaded
 region corresponds to the region that are excluded in our clustering
 analysis. In addition to the angular separation range of
 $20''<\theta<55''$, we exclude low-redshift pairs with 
$\theta \sim 10''$ \citep[see Eq.~1 of][for quantitative
details]{inada08} because most of the SQLS lens candidates in this
region were excluded based on the absence of any lensing objects in
the SDSS images \citep[see][]{inada08}, and therefore spectroscopic
follow-up observations for these candidates were highly incomplete.
Indeed, there is no binary quasar identified by the SQLS within the
exclusion region defined by Eq.~1 of \citet{inada08}, which supports
our argument above.

One of the most important issues for the clustering analysis is the
completeness and purity of our binary quasar sample. For instance, a
possible contamination in small-separation binary quasar samples 
comes from gravitational lensing.  However, the contamination of
gravitational lensing for our sample is expected to be negligibly
small by construction, because lensed quasars are the main focus of
the SQLS and hence very careful studies have been conducted for each
binary quasar to extensively check whether it is gravitational lensing
or binary quasar. For all the binary quasars with very similar
redshifts, deep follow-up images are obtained to look for any lensing
objects \citep[see][for details]{inada08,inada10,inada12}.

Perhaps more important is the completeness of our binary quasar
sample. One might argue that the candidate selection of the SQLS is
designed for lensed quasars and therefore may miss binary quasars with
large colour differences. However, the SQLS allows relatively large
colour differences for the lens candidate selection, considering the
effect of differential extinction that can significantly change colours 
of multiple images \citep[see][for details]{oguri06}. Also the
spectroscopic follow-up of the SQLS may not be complete in the sense
that many of the candidates have been rejected based on imaging
follow-up observations alone. We however note that the follow-up
strategy of the SQLS was such that any lens candidates that are very
likely quasar pairs based on the SDSS colour were almost always
followed-up spectroscopically in order to assure the high completeness
of the lens sample. To summarize, the completeness of the SQLS binary
quasar sample is expect to be very high, thanks to extensive
examinations of lens candidates conducted by the SQLS. Therefore,
throughout this paper, we simply assume that the binary quasar sample
is complete. This is in contrast to the small-scale clustering analysis 
by \citet{hennawi06} in which large corrections of the sample
incompleteness have been applied. We note that any incompleteness in our
binary quasar sample, if present, increases the clustering amplitude
further. This means that the clustering amplitudes presented in this
paper can also be interpreted as a strict lower limit of small-scale
clustering.  

\section{Clustering Analysis}
\label{sec:clusteringAnalysis}

\subsection{Measuring Correlation Function}
\label{sec:Wpmeasurement}
In order to reduce contamination by peculiar velocities of quasars,
we consider a projected two-point correlation function along the
line-of-sight. However, since the number of quasar pairs at
small-scale is very small, it is impractical to follow the common
procedure where we first measure two-dimensional (parallel and
perpendicular to the line-of-sight) correlation function and then
project it along the line-of-sight. Instead, we adopt an estimator
used by \citet{hennawi06}:  
\begin{equation}
 \bar{W}_p(r_{p, \rm min}, r_{p, \rm max})=\frac{\langle
  QQ\rangle}{\langle QR\rangle} -1,
\end{equation}
where $\langle QQ\rangle$ is the number of {\it companions} whose
transverse comoving separation $r_p$ falls in the range of $r_{p, \rm
rmin}<r_p<r_{p, \rm rmax}$.
For small-separation SQLS pairs, the number of companions is the
 same as that of pairs, because the counterparts that we search for
 are not included in the parent quasar catalogue. On the other
 hand, for large-separation DR7 pairs, the number of companions should
 be twice that of pairs because we consider the auto-correlation of
 the DR7 catalogue \citep[see also][]{hennawi06}.
The denominator $\langle QR\rangle$ is estimated as 
\begin{eqnarray}
 \langle QR\rangle
  &=&\sum_j^{N_{\rm DR7}} n(z_j, i<19.1)V_{\rm shell}^{\rm DR7}
(r_j, r_{p, \rm min}, r_{p, \rm max}) \nonumber\\
 &&\hspace*{-12mm}+\sum_j^{N_{\rm SQLS}} 
n(z_j, |i-i_j|<1.25)V_{\rm shell}^{\rm SQLS}(r_j, r_{p, \rm min}, 
r_{p, \rm max}),
\end{eqnarray}
where $N_{\rm DR7}$ and $N_{\rm SQLS}$ are the numbers of parent
catalogue of quasars from DR7 quasar catalogue ($N_{\rm DR7}=51,283$) 
and SQLS DR7 source quasar catalogue ($N_{\rm SQLS}=50,836$). The
small difference in the numbers of quasars is due to the removal of
SDSS fields with poor seeing conditions in the SQLS.
While $V_{\rm shell}^{\rm DR7}$ and $V_{\rm shell}^{\rm SQLS}$ are
basically a comoving volume of a cylindrical shell 
\begin{equation}
 V_{\rm shell}(z, r_{p, \rm min}, r_{p, \rm max})=\pi (r_{p, \rm
  max}^2-r_{p, \rm min}^2)\left[\frac{2v_{\rm max}}{H(z)/(1+z)}\right],
\end{equation}
we take account of the exclusion region shown in
Figure~\ref{fig:pairs} by evaluating $V_{\rm shell}^{\rm DR7}$ and 
$V_{\rm shell}^{\rm SQLS}$ only at larger and smaller scales than the
exclusion region, respectively. $H(z)$ is the Hubble parameter at
redshift of $z$, $n(z_j, i<19.1)$ is the number density of quasars
brighter than $i=19.1$ at the redshift of $z_j$, and 
$n(z_j, |i-i_j|<1.25)$ represents that of quasars whose magnitude
difference in $i$-band is smaller than 1.25 at redshift of $z_j$. 
We estimate the quasar number density in $i$-band using the $g$-band  
luminosity function of \citet{croom09}:
\begin{equation}
 \Phi(M_g, z)=\frac{\Phi^*}{10^{0.4(\alpha+1)(M_g-M_g^*(z))}+
10^{0.4(\beta+1)(M_g-M_g^*(z))}},
\end{equation}
\begin{equation}
 M_g^*(z)=M_g^*(0)-2.5(k_1z+k_2z^2),
\end{equation}
with the $K$-correction and the band conversion formulae provided by
\citet{richards06}
\footnote{Although the band conversion formula of
Eq. (2) in \citet{richards06} is for $z=0$, we apply the same formula
for $z=2$, considering that the conversion between $z=0$ and $z=2$ in a
same band can be performed using their Eq. (1) for each of $i$- and
$g$-band.}.
 The parameters are $(\alpha, \beta, \Phi^*, M_g^*(0), k_1,
k_2)=(-3.33, -1.42, 1.45\times 10^{-6} {\rm Mpc}^{-3}{\rm mag}^{-1},
-22.18, 1.44, -0.315)$.  We employ this $g$-band luminosity function to
reach deeper than $i=19.1$, which is needed to estimate $n(z_j,
|i-i_j|<1.25)$.  We confirmed that this model successfully reproduces
the number distribution of quasars as a function of redshift in the SDSS
DR7.  Because this luminosity function is
evaluated with cosmological parameters of $\Omega_m=0.3$,
$\Omega_\Lambda=0.7$ and $H_0=70{\rm \,km\,s^{-1}\,Mpc^{-1}}$, we use
these values whenever we convert the apparent magnitude to absolute
magnitude or correct the difference of comoving volume with under our
fiducial cosmological parameters.

\begin{figure}
 \includegraphics[width=84mm]{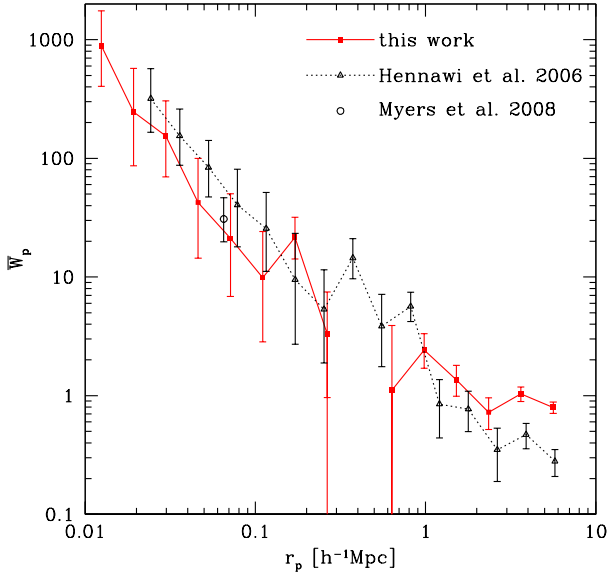}
 \caption{The measured projected correlation function $\bar{W}_p(r_p)$ 
   as a function of the comoving separation $r_p$, based on the binary
   quasar sample from the SQLS ({\it filled squares}). For comparison,
   previous small-scale clustering measurements by \citet{hennawi06}
   and \citet{myers08} are shown by filled triangles and open circle,
   respectively.}
 \label{fig:Wp}
\end{figure}

Figure~\ref{fig:Wp} plots the measured projected correlation function,
$\bar{W}_p$ with errors assuming the Poisson distribution.  Note that
the gap around $0.4 h^{-1}{\rm Mpc}$ is due to the exclusion of
$20''<\theta<55''$.  We find that the correlation function rises steeply
toward smaller scales down to $0.01h^{-1}{\rm Mpc}$, which basically
confirms the finding of \citet{hennawi06} that the small-scale
clustering amplitudes of quasars are high. However, the clustering
amplitudes in our analysis appears to be systematically smaller than
those of \citet{hennawi06}. We ascribe this difference to the
uncertainty associated with the completeness. We also plot the
clustering measurement by \citet{myers08}, who constructed a complete
binary quasar sample only for a narrow separation range to make a robust
clustering measurement. The result of \citet{myers08} is more consistent
with our result, which supports the argument that our binary quasar
sample is nearly complete.  
It is worth noting that there is a difference of $\bar{W}_p$ at
  $\theta>55''$ between our measurement and that of
  \citet{hennawi06}. While the origin of the difference is not clear, 
  we note that our measurement at $\theta>55''$ is in good agreement
  with clustering measurement of \citet{ross09}, as we will see
  later.

\subsection{Power-law Fit}
\label{sec:powerlaw}

We fit the measured correlation function with a single power-law of
the form $\xi(r)=(r/r_0)^{-\gamma}$, where $\xi(r)$ is the 
three-dimensional real-space correlation function. Such simple model 
is still very useful not only for quantifying the behavior of the
correlation function for a wide separation range, but also for
comparing correlation function measurements for different objects and
redshifts. 

Here we combine our small-scale clustering measurements with large-scale
measurements by \citet{ross09} to constrain the correlation function
parameters. More specifically, we simultaneously fit our measurements of
$\bar{W}_p(r_p)$ (Eq.~\ref{eq:WpPWfit}) at $r_p<0.3 h^{-1}{\rm Mpc}$ and
$w_p(r_p)$ (Eq.~\ref{eq:wpPWfit}) measurements by \citet{ross09} at
$r_p>0.3 h^{-1}{\rm Mpc}$ with two parameters, $r_0$ and $\gamma$. Note
that the fitting ranges of $r_p$ are chosen to avoid any overlap of
these two data. Although the redshift range of the \citet{ross09} quasar
sample, $0.3<z<2.2$, is slightly different from ours, we ignore this
small difference throughout the paper.  
Note that we adopt the Poisson error for $\bar{W}_p(r_p)$, whereas 
\citet{ross09} estimated the error on $w_p(r_p)$ using the jackknife
resampling. We ignore the difference because \citet{ross09} also
showed that their jackknife errors are not very 
different from the Poisson estimate at the scale considered here.
We summarize the calculation
of $\bar{W}_p(r_p)$ and $w_p(r_p)$ for the power-law model in
Appendix~\ref{sec:powerlawcalc}.

\begin{figure}
 \includegraphics[width=84mm]{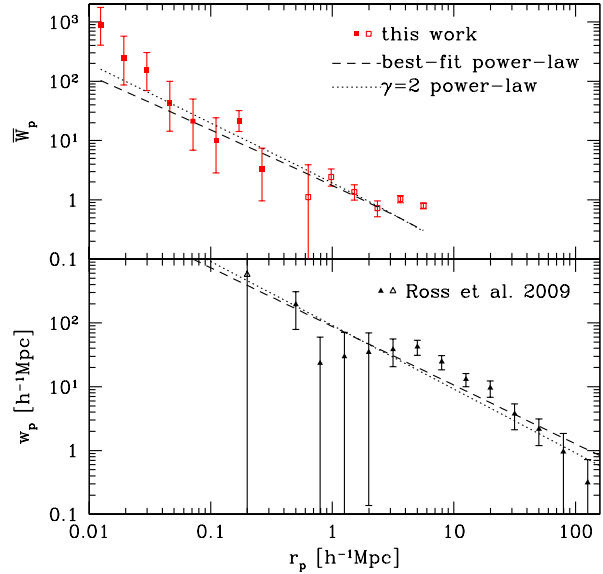}
 \caption{Power-law fit to the projected correlation functions. We show
 the result of the simultaneous fit to the projected correlation function
 $\bar{W}_p(r_p)$ ({\it upper panel}) measured in this paper and the
 large-scale correlation function $w_p(r_p)$ ({\it lower panel}) 
  measured by \citet{ross09}.  Dashed lines are the best-fit
 power-law model with two parameters $r_0$ and $\gamma$, while dotted
 lines are the result of power-law fitting with a fixed slope
 $\gamma=2$. The data points shown in filled symbols indicate that
 they are used for the fitting.} 
 \label{fig:WpwpPWfit}
\end{figure}

The comparison of the best-fit power-law model with the data is shown
in Figure~\ref{fig:WpwpPWfit}. We find the best-fit values to 
$r_0=5.37_{-0.39}^{+0.32} h^{-1}{\rm Mpc}$ and $\gamma=1.92_{-0.05}^{+0.04}$.  
These are quite consistent with the best-fit parameter values of
\citet{ross09}, $r_0=5.45_{-0.45}^{+0.35} h^{-1}{\rm Mpc}$ and
$\gamma=1.90_{-0.03}^{+0.04}$, which indicates that the measured
small-scale clustering is quite consistent with the extrapolation of
large-scale clustering measurements. Note that the reduced $\chi^2$ at
the best-fit values is 1.64, indicating that the measured correlation
function in fact deviates from a power-law, as is the case for the
correlation function of galaxies \citep{zehavi04}. We also consider
the case that the slope is fixed to $\gamma=2$, and find the best-fit 
correlation length for this case to $r_0=5.40_{-0.33}^{+0.31}
h^{-1}{\rm Mpc}$.

\section{Halo Model Interpretation}
\label{sec:halomodel}

\subsection{Model Description}

The Halo Occupation Distribution (HOD) framework
\citep{seljak00,peacock00,scoccimarro01} provides more physical
interpretation of the correlation function measurements. This model
assumes numbers of objects in individual dark haloes as a function of
halo masses, $\langle N(M)\rangle$, as well as how the objects are
distributed within haloes. The technique has been used for the 
analysis of the quasar/AGN correlation function measurements to study
quasar activities within a cosmological framework
\citep{porciani04,miyaji11,krumpe12,richardson12}.

In the halo model description, we divide the power spectrum into
two parts, the so-called one-halo and two-halo terms
\begin{equation}
 P(k)=P_{\rm 1h}(k) + P_{\rm 2h}(k).
\end{equation}
The projected correlation function can be obtained by \citep[e.g.,][]{cooray02}
\begin{equation}
 w_p(r_p)=\int_0^\infty \frac{k dk}{2\pi}P(k)J_0(kr_p),
  \label{eq:wppk}
\end{equation}
where $J_0(x)$ is the zeroth-order Bessel function of the first kind.

The two-halo term is simply estimated as
\begin{equation}
P_{\rm 2h}(k)=b^2 P_{\rm lin}(k),
\end{equation}
where $b$ is the linear bias parameter which can be modelled as
\begin{equation}
 b=\frac{1}{n_q}\int b_{\rm h}(M) \langle N(M)\rangle \frac{dn}{dM} dM,
\end{equation}
\begin{equation}
 n_q=\int \langle N(M)\rangle\frac{dn}{dM} dM,
\end{equation}
where we adopt the halo mass function $dn/dM$ and the halo bias
model $b_{\rm h}(M)$ of \citet{sheth99}. The linear matter power
spectrum $P_{\rm lin}(k)$ is computed using a fitting form of
\citet{eisenstein99} with the baryon density $\Omega_b=0.045$, the 
amplitude of fluctuations $\sigma_8=0.80$, and the slope of the initial
power spectrum $n_s=0.963$.  We model the mean number of quasars in
each halo, $\langle N(M)\rangle$, by the following simple form 
\begin{equation}
 \langle N(M)\rangle=f_N\times \frac{1}{\sqrt{2\pi}\Delta_m}
\exp\left[-\frac{\ln^2(M/M_m)}{2\Delta_m^2}\right].
  \label{eq:hod}
\end{equation}
This functional form is motivated by \citet{lidz06} who investigated
the luminosity dependence of quasar clustering and the quasar
luminosity functions and concluded that quasars with a broad
luminosity range reside in host haloes with a narrow range of masses. 
In addition, several observational works have shown that there is a
deficit of quasar lying in massive clusters, at least at $z\la
1$ \citep[e.g.,][]{green11,farina11,harris12}, which also supports
this HOD model. Our model is also consistent with a recent HOD
analysis result of X-ray AGNs by \citet{miyaji11}. Following the
argument of \citet{lidz06}, in this paper we fix $\Delta_m=0.75$,
leading to a fairly narrow mass range of $\sim 2$~dex, although we
confirmed that our conclusion is insensitive to the choice of
$\Delta_m$. The normalization of Eq.~(\ref{eq:hod}), $f_N$, is
calculated for each HOD parameter set by matching the number density
of quasars to the observed one,  $n_q \sim 1.35\times 10^{-6} h^3 {\rm
  Mpc}^{-3}$ at the mean redshift of $z=1.40$ for our sample.

The one-halo term can be described as \citep{seljak00}
\begin{equation}
 P_{\rm 1h}(k)=\frac{1}{n_q^2}\int 
  \langle N(N-1)\rangle u(k, M)^p
  \frac{dn}{dM}dM.
\end{equation}
Here, $u(k, M)$ is the Fourier transform of the number density profile
of quasars in a dark matter halo with mass $M$ which is normalized
to unity at $k\to 0$. The parameter $p$ is basically determined by the
fraction of quasars that reside in the centre of the halo. 
In this paper, we consider a simple model that assumes a constant
``satellite fraction'' $f_{\rm sat}$. This parameter is defined by the
fraction of quasars that appears as satellites in dark matter haloes.
Then the HOD of central and satellite components \citep[see,
  e.g.,][]{berlind02,kravtsov04,zheng05} are described by 
$\langle N_{\rm cen}\rangle = (1-f_{\rm sat})\langle N(M)\rangle$ 
and $\langle N_{\rm sat}\rangle = f_{\rm sat}\langle N(M)\rangle$,
respectively.
We ignore the mass dependence of $f_{\rm sat}$, which is easily
justified when we consider relatively narrow range of halo masses as
assumed in this paper. 
Furthermore, we assume that the central quasar activity
and the satellite activity are independent of each other.
In this case, we can simplify the integral kernel as 
\begin{eqnarray}
 \langle N(N-1)\rangle u(k, M)^p && \nonumber\\
&&\hspace*{-32mm}\equiv 2\langle N_{\rm cen}N_{\rm
  sat}\rangle u(k, M) +\langle
  N_{\rm sat}(N_{\rm sat}-1)\rangle |u(k, M)|^2 \nonumber\\
&&\hspace*{-32mm}=\left[2f_{\rm sat}(1-f_{\rm sat})u(k, M)
+f_{\rm sat}^2|u(k, M)|^2\right]\langle N(M)\rangle^2,
\label{eq:NN}
\end{eqnarray}
where we assumed Poisson statistics. In this model, neither the one-halo
nor two-halo term depend on the normalization $f_N$ in Eq.~(\ref{eq:hod}).

For the number density profile of quasars in a dark matter halo, we
employ the NFW profile \citep{nfw} with the concentration parameter
$c_{\rm vir}$ of \citet{bullock01} or a Singular Isothermal Sphere
(SIS) profile. Both the profiles are truncated at the virial radius of
each dark matter halo. 

\subsection{Data Conversion: $\bar{W}_p$ to $w_p$}
\label{sec:wpconv}

While our measurement of $\bar{W}_p(r_p)$ enables direct
 comparisons with \citet{hennawi06}, it is computationally more 
 efficient to work on $w_p(r_p)$ for the comparison of the HOD model
 with the data. Here we attempt to convert our measurements of
 $\bar{W}_p(r_p)$ to $w_p(r_p)$ adopting the following
 approximation. The conversion makes it easier to compare our results
 with other correlation function measurements.

We do so by multiplying the following approximated conversion 
factor to $\bar{W}_p(r_p)$
\begin{equation}
 \int_{z_{\rm min}}^{z_{\rm max}}dz \frac{dV_c}{dz}n(z) 2v_z\times
  \frac{\bar{w}_p(r_p)}{\bar{w}_p(r_p, v_{\rm max})},
  \label{eq:Wpwpfactor}
\end{equation}
where
\begin{eqnarray}
 \bar{w}_p(r_p, v_{\rm max})&\equiv& 
  \frac{1}{N}\int_{z_{\rm min}}^{z_{\rm max}} dz\frac{dV_c}{dz} n(z)\nonumber\\
  &&\times \int_0^{v_z} d\pi\xi(\sqrt{r_p^2+\pi^2}, z),
\end{eqnarray}
and $\bar{w}_p(r_p)\equiv \bar{w}_p(r_p, v_{\rm max}\to\infty)$.  The
second term of Eq~(\ref{eq:Wpwpfactor}) is evaluated assuming that the
real-space correlation function is a power-law with $r_0=5.37 
h^{-1}{\rm Mpc}$ and $\gamma=1.92$.  

\begin{table}
 \caption{Projected correlation functions $\bar{W}_p(r_p)$ and
   $w_p(r_p)$ measured in this paper. Note that $w_p(r_p)$ in this
   table are approximated values computed by the method described in
   \S\ref{sec:wpconv}. Values in parentheses indicate 
   measurement errors. \label{tab:wp}}   
 \begin{tabular}{@{}ccc}
  \hline
   $r_p$ 
   & $\bar{W}_p(r_p)$
   & $w_p(r_p)$ \\
   ($h^{-1}{\rm Mpc}$)
   &
   & ($h^{-1}{\rm Mpc}$) \\
 \hline
  0.012 & 888.748 (513.696)&  42157.4 (24367.0)\\
  0.019 & 246.645 (175.112)&  11703.3 (8309.1)\\
  0.030 & 154.087 (89.539) &  7315.0 (4250.7)\\
  0.046 &  42.409 (30.695) &  2014.7 (1458.2)\\
  0.071 &  21.161 (15.671) &  1006.4(745.3)\\
  0.110 &   9.850 (7.672)  &  469.2 (365.5)\\
  0.171 &  21.545 (7.515)  &  1028.8 (358.9)\\
  0.265 &   3.304 (2.485)  &  158.3( 119.1)\\
  0.409 &  $\cdots$        &   $\cdots$ \\
  0.634 &   1.114 (1.495)  &   54.2 (72.7)\\
  0.981 &   2.429 (0.731)  &  119.5 (36.0)\\
  1.518 &   1.361 (0.373)  &   68.2 (18.7)\\
  2.349 &   0.725 (0.206)  &   37.4 (10.6)\\
  3.636 &   1.037 (0.145)  &   55.9 (7.8)\\
  5.627 &   0.795 (0.088)  &   45.9 (5.1)\\
 \hline
 \end{tabular}
\end{table}

\begin{figure}
 \includegraphics[width=84mm]{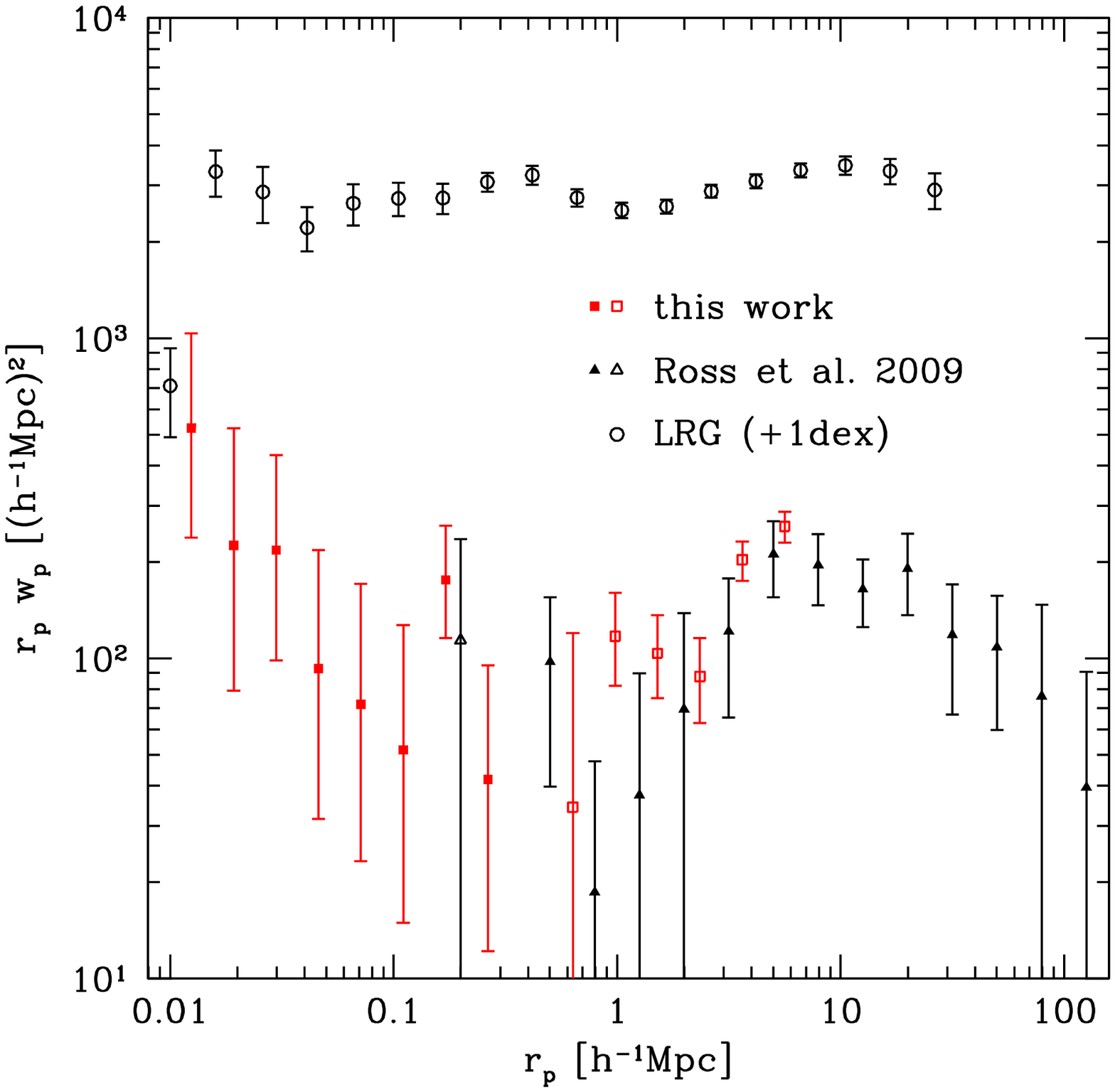}
 \caption{Projected correlation function $w_p(r_p)$ of quasars from
   our measurement ({\it squares}) and and that of \citet{ross09}
   ({\it triangles}). For comparison, we also plot the projected
   correlation function of LRGs at $z\sim 0.3$ ({\it open circles}) 
   measured by \citet{masjedi06} and \citet{zehavi05}. Note that the
   LRG correlation function is shifted upward by 1~dex for
   illustrative purpose.
 The data points shown in filled symbols indicate that
 they are used for the fitting.}
 \label{fig:wp}
\end{figure}

We show the estimated $w_p(r_p)$ as well as the original projected
correlation function in Table~\ref{tab:wp}. We also plot $w_p(r_p)$
from our measurements and that from \citet{ross09} in Figure~\ref{fig:wp}.
For comparison, we also show the correlation function of Luminous Red
Galaxies (LRGs) at $z\sim 0.3$ measured by \citet{masjedi06} and
\citet{zehavi05}.  We find that the overall shape of $w_p(r_p)$ for
quasars is quite similar to that for LRGs. Both the correlation
functions show deviations from a power-law in a similar manner, which
can be better fitted by a more general HOD model as we will see in
\S\ref{sec:hodresult}. We however note a possible difference in the
shapes of $w_p(r_p)$ at the smallest separations of $r_p\sim
10h^{-1}{\rm kpc}$. The LRG correlation function appears to decrease
significantly at the smallest bin, which is also suggested by the
recent measurement of the satellite galaxy distribution around LRGs
\citep{tal12}, whereas there is a hint of excess at the smallest
separations in the quasar correlation function. The difference might
be explained by different effects of direct interactions on quasars
and galaxies. 

\subsection{Result of Fitting}
\label{sec:hodresult}

\begin{figure}
 \includegraphics[width=84mm]{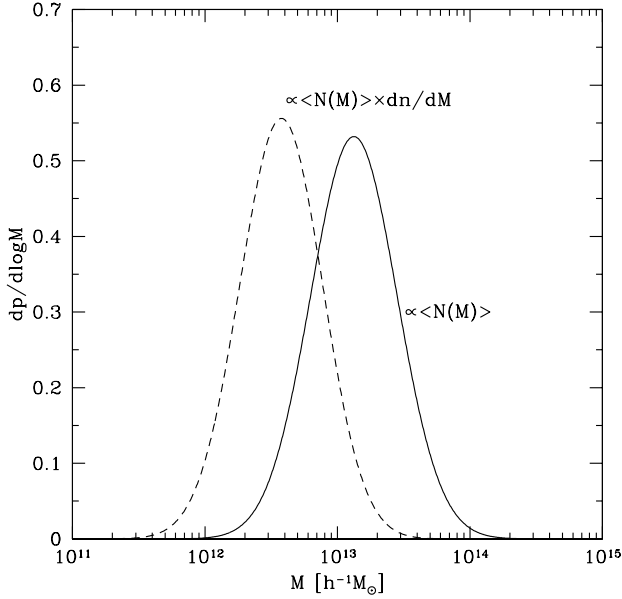}
 \caption{
   The best-fit HOD model $\langle N(M)\rangle$ ({\it solid})
   and the host halo mass distribution $\langle N(M)\rangle dn/dM$
   ({\it dashed}). Both curves are normalized so that it becomes unity
   when integrated over $d\log M$. 
 }  \label{fig:nhalo}
\end{figure}

Here we compare the HOD model predictions with measurements of
$w_p(r_p)$ derived in \S\ref{sec:wpconv}. For simplicity, calculations
of $w_p(r_p)$ with the HOD model are done at the mean redshift of our
parent quasar samples of $z=1.40$. Our HOD model involves only two
parameters, $M_m$ and $f_{\rm sat}$. Considering the fact that the
two-halo term is independent of $f_{\rm sat}$ and that the data points
of \citet{ross09} lie in the scale where the two-halo term is dominant,
we first use the measurement of \citet{ross09} at $r_p>0.3 h^{-1}{\rm
Mpc}$ to fit $M_m$ only. The best-fit value is found to
$M_m=1.33_{-0.32}^{+0.36}\times 10^{13} h^{-1}M_\odot$, and the
resulting normalization is $f_N=7.6\times 10^{-3}$.  
This means that dark matter halos with the peak mass of
$M=M_m$ harbour $\sim 4\times 10^{-3}$ quasars on the average.
Figure~\ref{fig:nhalo} plots the best-fit $\langle N(M)\rangle$ and
the host halo mass distribution $\langle N(M)\rangle dn/dM$. The
latter which indicates the typical mass of host halo harboring
quasars peaks at $\sim 3\times 10^{12} h^{-1}M_\odot$, which is
consistent with previous measurements for optical quasars
\citep[e.g.,][]{shen07,coil07,ross09,richardson12}. 
Next, we fit the small-scale
measurements derived in this paper to constrain $f_{\rm sat}$ with a
fixed peak mass in $\langle N(M)\rangle$, $M_m=1.33\times
10^{13}h^{-1}M_\odot$.

\begin{figure}
 \includegraphics[width=84mm]{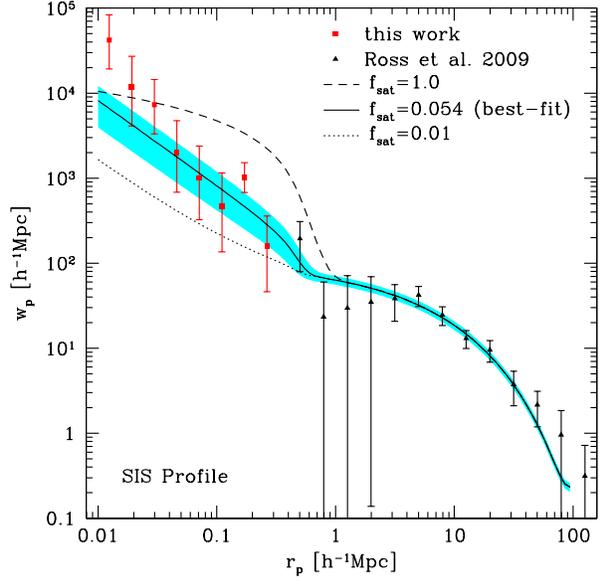}
 \caption{HOD model predictions of the projected correlation function
   $w_p(r_p)$ for three different satellite fraction, 
   $f_{\rm sat}=1$ ({\it dashed}), $0.054$ ({\it solid}), and $0.01$
   ({\it dotted}). An
   SIS profile is assumed for the radial number density profile of
   satellites. The peak mass in $\langle N(M)\rangle$, $M_m$ in
   Eq.~(\ref{eq:hod}), is fixed to the 
 best-fit value. 
  The shaded region around the best-fit model represents $1\sigma$
  range.
  }  \label{fig:fsat} 
\end{figure}

We find that $f_{\rm sat}=0.054_{-0.016}^{+0.017}$ for an SIS profile
and $f_{\rm sat}=0.048_{-0.015}^{+0.016}$ for an NFW profile.  Combining
the $\chi^2$ of the large-scale and small-scale fitting, the reduced
$\chi^2$s for the best-fit parameters are 0.83 for the SIS profile and
0.89 for the NFW profile, which are significantly better than the
reduced $\chi^2$ of 1.64 for the power-law fit
shown in \S\ref{sec:powerlaw}, despite the same number
of free parameters these models contain.  Figure~\ref{fig:fsat}
illustrates how the small-scale clustering measurements presented in
this paper constrain $f_{\rm sat}$.  Basically $f_{\rm sat}$ controls
the amplitude of the one-halo term such that larger satellite fractions
predict larger small-scale clustering amplitudes.

\begin{figure}
 \includegraphics[width=84mm]{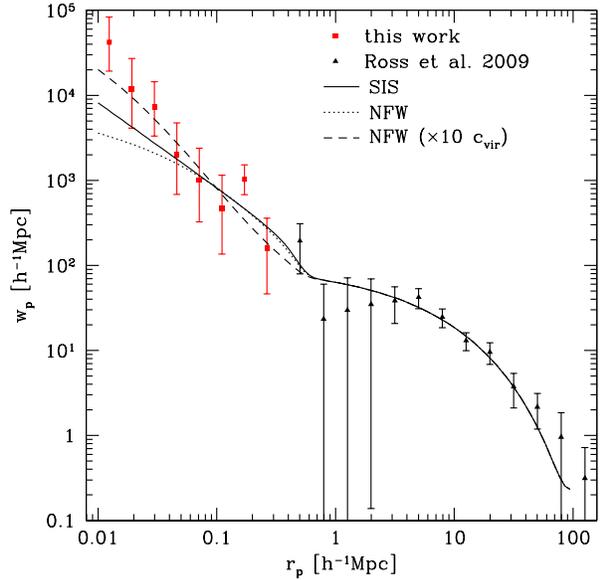}
 \caption{Comparisons of the best-fit HOD model predictions for the
   projected correlation function assuming SIS ({\it solid}) and NFW
   ({\it dotted}) profiles with the measurements. The dashed line
   shows the HOD model assuming an NFW profile, but with 10 times
   larger concentration parameter than the fiducial \citet{bullock01}
   model. }
 \label{fig:profile}
\end{figure}

Figure~\ref{fig:profile} compares our best-fit HOD models for SIS and
NFW profiles with the measurements. Except for a few points at the
smallest scales, our HOD model reproduces the global feature of the
measured correlation function very well. At the smallest scale around 
$\sim 10 h^{-1}{\rm kpc}$, the difference of assumed number
density profile of quasars in a halo becomes prominent as expected,
although even the steep SIS profile fails to reproduce the clustering
amplitude at these scales. Hence, we also consider the case of an NFW
profile with the concentration parameter value enhanced by a factor of
10 compared with our fiducial model \citep{bullock01}. This enhanced
concentration model predicts $c_{\rm vir}\sim 30$ for haloes with mass
$M=M_m$ at $z=1.4$.  The result shown in Figure~\ref{fig:profile}
indicates that this high-concentration model better reproduces the 
strong small-scale clustering found in the measurements. 

\section{Conclusion}
\label{sec:summary}

We have measured the small-scale correlation function of low-redshift
($0.6<z<2.2$) quasars using a new sample of 26 spectroscopically
confirmed binary quasars from the SQLS. The angular separation range of
$1''<\theta<20''$ for the binary quasar sample allows us to measure the
quasar auto-correlation function from the projected comoving separation
$r_p\sim 200h^{-1}{\rm kpc}$ down to $r_p\sim 10h^{-1}{\rm kpc}$. To do
so, we have assumed that our binary quasar sample is complete within a
specified range of redshifts and angular separations, which is based on
careful selections and extensive follow-up observations of the SQLS in
order to construct a complete gravitational lens sample. We have found
that the correlation function rises steeply toward smaller
separations. Our result is in reasonable agreement with earlier result
by \citet{hennawi06}, in which large corrections of sample
incompleteness has been applied to measure small-scale clustering
signals. While the measured small-scale clustering in our analysis
appears to be slightly smaller than the result of \citet{hennawi06}, we
note that our result is more consistent with the clustering amplitude
measurement within a narrow separation range presented by
\citet{myers08} based on a complete binary quasar sample.  By combining
our measurements with the large-scale correlation function obtained by
\citet{ross09}, we have fitted the correlation functions for a wide
range of $0.01h^{-1}{\rm Mpc}\la r_p \la 10h^{-1}{\rm Mpc}$ by a
power-law $\xi(r)=(r/r_0)^{-\gamma}$, and found the best-fit values to
$r_0=5.37_{-0.39}^{+0.32} h^{-1}{\rm Mpc}$ and
$\gamma=1.92_{-0.05}^{+0.04}$. The measured correlation function can be
described reasonably well by this single power-law model over 4 order of
magnitudes in $r_p$.

We have interpreted the measured quasar correlation function within the
framework of the HOD. In particular, we have proposed a simple HOD
model that assumes a constant satellite fraction $f_{\rm sat}$, which
is a reasonable assumption when the average number of quasars in
haloes with mass $M$, $\langle N(M)\rangle$, becomes large within
relatively narrow range in masses. Using this model, we have shown
how the small-scale clustering can be used to constrain $f_{\rm
  sat}$. The best-fit satellite fraction of our quasar sample is
$f_{\rm sat}=0.054_{-0.016}^{+0.017}$ for an SIS profile and 
$f_{\rm sat}=0.048_{-0.015}^{+0.016}$ for an NFW profile. The HOD
model reproduces the measured correlation function quite well, except
at the smallest separations of $r_p\sim 10h^{-1}{\rm 
kpc}$ where the model slightly under-predicts clustering signals even if
we assume the steep SIS profile for the spatial distribution of
satellite components. We find that even steeper profiles, such as an
NFW profile with the concentration parameter of $c_{\rm vir}\ga 30$,
are needed to explain the measured clustering amplitudes at the
smallest scale. 
Recent numerical simulations of low-luminosity AGN also suggest that
radial profile of AGN inside haloes is much steeper than the normal NFW
profile \citep{degraf11, chatterjee12}.
Such excess clustering at the smallest separations,
as well as possible different behaviors between quasar and LRG
correlation functions at those scales
 \citep[see][for a discussion on the radial profile of LRGs from HOD
 modelling]{watson10},
might be due to an enhanced
quasar activity by mergers. 
Direct high-resolution imaging of host
galaxies for these very small-separation binary quasars might be
helpful for testing this hypothesis.

When we almost finished this paper, we found that \citet{richardson12}
conducted similar HOD analysis of quasar clustering to constrain the
satellite fraction, using small-scale clustering measurements of
\citet{hennawi06}. However, their best-fit satellite fraction of $f_{\rm
sat}=(7.4\pm1.4)\times 10^{-4}$ for quasars at $z\sim 1.4$ appear to be
significantly smaller than our result.  This discrepancy is explained by
different forms of HOD modelling adopted in both papers. Their $\langle
N(M)\rangle$ extends toward the mass scale of massive clusters where
satellite components dominate the mean number of quasars, whereas in our
HOD model we have assumed a deficit of quasars in very massive clusters.
We confirmed that their best-fit HOD model indeed predicts significantly
smaller $f_{\rm sat}$ than our result.  Also note that we have assumed
that the quasar activities of the central and the satellite quasars are
independent of each other, which means that quasars can be
satellites even if the central object is not a quasar. When there exists a
strong central-satellite correlation such that haloes are allowed to host
satellite quasars only if the central galaxy is also a quasar, we expect
that smaller $f_{\rm sat}$ can recover the small-scale clustering of
quasar pairs. In this sense, our best-fit $f_{\rm sat}$ value may be
considered as an upper limit. Detailed observation of the environment
of (binary) quasars or dynamical analysis of quasar pairs might provide
more direct information on the mass of host haloes of quasars and on the
location of quasars in the host haloes
\citep[e.g.,][]{fukugita04,serber06,boris07,strand08,green11,farina11,harris12}.

\section*{Acknowledgments}
We thank an anonymous referee for many useful suggestions.
This work was supported in part by the FIRST program ``Subaru
Measurements of Images and Redshifts (SuMIRe)'', World Premier
International Research Center Initiative (WPI Initiative), MEXT, 
Japan, and Grant-in-Aid for Scientific Research from the JSPS
(23740161, 24740171). 


\appendix
\section{Power-Law Model Calculation}
\label{sec:powerlawcalc}

We derive the relations between the power-law model of
$\xi(r)=(r/r_0)^{-\gamma}$ and the projected correlation functions
$\bar{W}_p(r_p)$ and $w_p(r_p)$.

The projected correlation function, $\bar{W}_p$, can be interpreted as
an average over the parent quasar sample
\begin{equation}
 \bar{W}_p(r_{p, \rm min}, r_{p, \rm max})=
  \frac{1}{N_{\rm QSO}}\sum_j^{N_{\rm QSO}}
\bar{W}_p(z_j, r_{p, \rm min}, r_{p, \rm max}),
\end{equation}
where $\bar{W}_p(z_j, r_{p, \rm min}, r_{p, \rm max})$ is computed as
\begin{equation}
 \bar{W}_p(z, r_{p, \rm min}, r_{p, \rm
  max})=\frac{\int_{-v_z}^{v_z}
  d\pi\int_{r_{p, \rm min}}^{r_{p, \rm max}}dr_p 2\pi r_p 
 \xi(r_p, \pi, z)}{V_{\rm
  shell}(z, r_{p, \rm min}, r_{p, \rm max})},
\end{equation}
with $\xi(r_p, \pi)$ being the anisotropic redshift-space correlation
function and $v_z\equiv v_{\rm max}(1+z)/H(z)$ \citep{hennawi06}.
In this paper, we employ the following approximated expressions in
order to speed up the computation:
\begin{equation}
 \bar{W}_p(r_p)\sim\frac{1}{N}\int_{z_{\rm min}}^{z_{\rm max}}
 dz\frac{dV_c}{dz} n(z)
  \frac{1}{v_z}\int_0^{v_z} d\pi\xi(\sqrt{r_p^2+\pi^2}, z),
  \label{eq:Wpapprox}
\end{equation}
where $\xi(r, z)$ is the isotropic real-space correlation function, 
$z_{\rm min}=0.6$ and $z_{\rm max}=2.2$ are the minimum and maximum of
the redshift range, $r_p$ is set to $(r_{p, \rm min}+r_{p, \rm
  max})/2$, $n(z)$ is the comoving 
number density of quasars (brighter than $i=19.1$ for our case), and
\begin{equation}
 N=\int_{z_{\rm min}}^{z_{\rm max}} dz\frac{dV_c}{dz} n(z).
\end{equation}
When $\xi(r)$ is described by a single power-law of
$\xi(r)=(r/r_0)^{-\gamma}$, Eq~(\ref{eq:Wpapprox}) can be 
simplified as
\begin{eqnarray}
 \bar{W}_p(r_p)&\sim&\frac{1}{N}\int_{z_{\rm min}}^{z_{\rm max}}
 dz\frac{dV_c}{dz} n(z)\nonumber\\
&&\times
  \left(\frac{r_0}{r_p}\right)^\gamma
  \ _2F_1\left(\frac12, \frac{\gamma}{2}, \frac32,
	  -\frac{v_z^2}{r_p^2}\right),
  \label{eq:WpPWfit}
\end{eqnarray}
where $_2F_1(a,b,c;x)$ is the Gauss' hypergeometric
function\footnote{For numerical calculation of the Gauss'
hypergeometric function, see \citet{michel08}.}.

On the other hand, the projected correlation function $w_p(r_p)$,
which is more commonly used, is defined by
\begin{equation}
 w_p(r_p)=2\int_0^\infty d\pi \xi(\sqrt{r_p^2+\pi^2}).
\end{equation}
Again, when $\xi(r)$ is described by a single power-law of
$\xi(r)=(r/r_0)^{-\gamma}$ with $\gamma>1$, $w_p(r_p)$ can be computed
as
\begin{equation}
 w_p(r_p)=\frac{r_0^\gamma
  \sqrt{\pi}\Gamma((\gamma-1)/2)}{\Gamma(\gamma/2)}r_p^{1-\gamma},
  \label{eq:wpPWfit}
\end{equation}
where $\Gamma(x)$ is the Gamma function.

\label{lastpage}

\end{document}